\documentclass[aps,prd,showpacs,nofootinbib,twocolumn,floatfix,superscriptaddress,preprintnumbers]{revtex4}
\usepackage[dvips]{graphicx}

\def\10{$SO(10)$}
\def\21{SU(2) $\otimes$ U(1) }

\def\422{$SU(4) \otimes SU(2) \otimes SU(2)$}
\def\321{SU(3) $\otimes$ SU(2) $\otimes$ U(1)}

\newcommand {\ignore}[1]{}

\def\lsim{\raise0.3ex\hbox{$\;<$\kern-0.75em\raise-1.1ex\hbox{$\sim\;$}}}
\def\gsim{\raise0.3ex\hbox{$\;>$\kern-0.75em\raise-1.1ex\hbox{$\sim\;$}}}

 \newcommand{\ba}{\begin{array}}
\newcommand{\ea}{\end{array}}
\relax
\def\321{$SU(3)\times SU(2)\times U(1)$}

\begin{document}
\renewcommand{\Huge}{\Large}
\renewcommand{\LARGE}{\Large}
\renewcommand{\Large}{\large}

\title{Optical activity of relic neutrino-antineutrino gas }
\author{V. B. Semikoz} \email{semikoz@yandex.ru}
\affiliation{
 IZMIRAN,  Kaluzhskoe highway, 4, Moscow, Troitsk, 142190, Russia}

\date{\today}
\begin{abstract}
We revise a problem of the birefringence of electromagnetic waves in a chiral medium in the Standard Model (SM) of particle physics
arising in an isotropic plasma due to parity violation. The recent calculations of the weak correction to the photon polarization operator
in electroweak plasma allow significantly to improve some previous estimates of such effect in astrophysics. Nevertheless, it has remained beyond the abilities of the present technics yet.
\end{abstract}


\maketitle

{\bf Maxwell equations in a chiral isotropic medium}

In an isotropic plasma accounting for electroweak interactions in the Standard Model (SM) Maxwell equations in the Fourier  representation take the form  \cite{Nieves:1992et}:
\begin{eqnarray}\label{Maxwell} 
&&{\bf k}\cdot{\bf B}=0,~~~{\bf k}\times{\bf E}=\omega {\bf B},~~ i\epsilon{\bf k}\cdot{\bf E}=\rho_{ext},\nonumber\\&&
\mu^{-1}i{\bf k}\times {\bf B} + i\omega \epsilon {\bf E} + i\zeta\omega {\bf B}={\bf j}_{ext},
\end{eqnarray}
where $\epsilon$ is the dielectric permittivity, $\mu$ is the magnetic permeability, $\zeta$ is the chiral dispersion characteristic in plasma. In SM for an isotropic medium this third dimensionless constant describing the electromagnetic properties of plasma (in addition to the two standard ones, $\epsilon$, $\mu$) should obey some C,P,T symmetry properties. The requirement that the fields ${\bf E}$, ${\bf B}$ and the current density ${\bf j}$ are real quantities in coordinate space plus  the time reversal symmetry $\zeta (k, - \omega)= - \zeta (k, \omega)$ imply that $\zeta$ is pure imagine and odd function of $\omega$ \cite{Nieves:1992et}. Note that in isotropic plasma one can put $\mu=1$ since in the high-frequency limit $\omega\gg k<v>$ one can neglect spatial dispersion. As a result the general permittivity tensor $\epsilon_{ij}(\omega,k)=(\delta_{ij} - k_ik_j/k^2)\epsilon^{tr}(\omega,k) + \epsilon^l(\omega,k)k_ik_j/k^2$ for $\epsilon^{tr}(\omega)=\epsilon^l(\omega)=1 - \omega_p^2/\omega^2=\epsilon (\omega)$ takes the
trivial form $\epsilon_{ij}=\delta_{ij}\epsilon (\omega)$, where $\omega_p^2=e^2n_e/m_e$ is the plasma frequency. Then the well-known relation 
\begin{equation}\label{permeability}
1 - \frac{1}{\mu (\omega, k)}=\frac{\omega^2}{k^2}\left[\epsilon^{tr}(\omega, k) - \epsilon^{l}(\omega,k)\right]\to 0
\end{equation}
gives $\mu=1$. Thus, we use below the two dispersion characteristics $\epsilon (\omega)$ and $\zeta (\omega)$ obeying known symmetry properties. The problem is how to find a new chiral parameter $\zeta (\omega)$ in SM. Note that parity conservation in QED plasma automatically gives $\zeta=0$ while extending model to SM one can expect $\zeta\neq 0$. In plasma with {\it different chemical potentials} of left and right fermions, $\mu_L\neq \mu_R$, there appears a parity-odd term in polarization operator of photons through the one-loop weak correction to the QED polarization operator
 \cite{Boyarsky:2012ex},
\begin{equation}\label{polarization}
\Pi_2 (0)=\frac{\alpha_{em}}{2\pi}\frac{4G_F}{\sqrt{2}}\left[c_{L_{\alpha}}L_{\alpha} + c_BB\right],
\end{equation}
where coefficients $c_{L_{\alpha}}$, $c_B\sim O(1)$ depend on the fermionic content of the plasma; $L_{\alpha}$ and $B$ are lepton and baryon asymmetries correspondingly; $G_F=10^{-5}/m_p^2$ is the Fermi constant; $\alpha_{em}=e^2/4\pi=1/137$ is the fine-structure constant. 
The polarization operator (\ref{polarization}) arises as the factor $X_0$ in the Chern-Simons (CS) anomaly term  entering the effective Lagrangian density, $\propto\varepsilon_{\alpha\beta\mu\nu}X_{\alpha}A_{\beta}\partial_{\mu}A_{\nu}$, being caused by parity violation in SM. While the difference $\mu_L - \mu_R$ vanishes at the tree level the weak corrections to chemical potentials $\delta\mu_L - \delta\mu_R\neq 0$ induced by Fermi interactions provide a non-zero $\Pi_2(0)\propto \sum_f q_f^2(\delta\mu_{f_L} - \delta\mu_{f_R})$ with the sum over all conserved fermion charges \cite{Boyarsky:2012ex}.

In the absence of external currents and charges, $\rho_{ext}={\bf j}_{ext}=0$, accounting for the additional pseudovector current ${\bf j}_{corr}=\Pi_2(0){\bf B}(\omega,{\bf k})$ given by the CS term we get a modified Maxwell equation generalized in SM due to parity violation: 
\begin{equation}\label{Maxwell3}
 i\omega {\bf E}(\omega,{\bf k}) + i{\bf k}\times {\bf B}(\omega,{\bf k})={\bf j}_{ind}(\omega,{\bf k})+ \Pi_2 (0){\bf B}(\omega,{\bf k}) .
\end{equation} 
In the right hand side (\ref{Maxwell3}) the induced vector current ${\bf j}_{ind}(\omega,{\bf k})=\sigma_{cond}(\omega){\bf E}(\omega,{\bf k})=-i(\epsilon(\omega) - 1)\omega{\bf E}(\omega,{\bf k})$ is the standard ohmic current in correspondence with the relation of dielectric permittivity and conductivity in isotropic plasma, $\epsilon (\omega)=1 +i\sigma_{cond}(\omega)/\omega$. Substituting this vector current into Eq. (\ref{Maxwell3}) one can recast the second line in (\ref{Maxwell}) as
\begin{equation}\label{Maxwell4}
i{\bf k}\times {\bf B}(\omega,{\bf k}) + i\omega \epsilon (\omega){\bf E}(\omega,{\bf k}) + i\zeta(\omega)\omega {\bf B}(\omega,{\bf k})= 0,
\end{equation}
where the third dispersion characteristic $\zeta$, \begin{equation}\label{zeta}
\zeta(\omega)=\frac{i\Pi_2(0)}{\omega},
\end{equation} takes the explicit form being pure imagine and odd function of $\omega$ as it should be.

{\bf Birefringence of electromagnetic waves in a chiral isotropic plasma}

Substituting ${\bf B}=({\bf k}\times {\bf E})/\omega$ one can easily get from (\ref{Maxwell4}) the dispersion equation for a right-circular and left-circular states of electromagnetic transversal waves,\\ ${\bf k}\cdot{\bf E}=0$, ${\bf E}={\rm E}(\omega,k)\hat{e}_{\pm}$, 
\begin{equation}\label{dispersion}
\omega^2 -k^2\left(\frac{1}{\epsilon} \pm \frac{i\omega\zeta}{\epsilon k}\right)=0,
\end{equation}
where $$
\hat{e}_{\pm}=\frac{1}{\sqrt{2}}({\bf e}_1 \pm i{\bf e}_2), ~~{\bf e}_2={\bf k}\times{\bf e}_1/k, 
$$
are the right and left polarization vectors, ${\rm E}(\omega,k)$ is the wave amplitude. Substituting weak chiral parameter (\ref{zeta}), $\Pi_2(0)\ll \omega,~k$, one obtains from the dispersion equation (\ref{dispersion}) the two dispersion relations $\omega_{\pm}(k)$ for a fixed wave number $k$:
\begin{equation}\label{frequency}
\omega_{\pm}=\sqrt{\omega_p^2 + k^2} \mp \left(\frac{\Pi_2(0)}{2}\right)\frac{k}{\sqrt{\omega^2_p + k^2}},
\end{equation}
or the two wave numbers for a fixed frequency $\omega$:
\begin{equation}\label{wavenumber}
k_{\pm}= \sqrt{\omega^2 - \omega_p^2} \pm \frac{\Pi_2(0)}{2}.
\end{equation}
One of the issues of optical activity in media is the rotation of polarization vector in the plane perpendicular to the direction of wave propagation. Choosing  z-axis parallel to the photon momentum, ${\bf k}=(0,0,k)$, we can treat at the source position (point $z=0$) a plane polarized wave as equal admixture of a right-circular and a left-circular polarized waves,
\begin{eqnarray}\label{admixture}
&&{\bf E}(z,t)=E_{\omega}e^{-i\omega t}\left(e^{ik_+z}\hat{e}_+ + e^{ik_-z}\hat{e}_-\right)=E_{\omega}e^{-i\omega t}\times\nonumber\\
&&\times e^{i(k_+ +k_-)z/2}\left(e^{i(k_+ -k_-)z/2}\hat{e}_+ + e^{-i(k_+ -k_-)z/2}\hat{e}_-\right).\nonumber\\
\end{eqnarray}
Here at the point $z=0$ (source position) the polarization vector is directed along ${\bf e}_1$, chosen as $x$-axis. Then it rotates in $x,y$-plane and points at the distance $z=l$ at the angle (relative to the $x$-axis) given by 
\begin{equation}\label{angle}
\phi (l)=\frac{1}{2}\left(k_+ - k_-\right)l=\Pi_2(0)l.
\end{equation}
Let us discuss some applications and compare predictions of different models of chiral media.

{\bf Rotary power in models \cite{Mohanty:1997mr,Abbasabadi:2001ps,Abbasabadi:2003uc} and present work}

It is instructive to compare the results of different calculations for the rotary power $\Phi=\phi/l$. For a photon $k_{\mu}=(\omega,{\bf k})$ propagating in vacuum ($k_{\mu}k^{\mu}=0$) filled by the neutrino-antineutrino sea one obtains \cite{Abbasabadi:2001ps,Abbasabadi:2003uc};
\begin{equation}\label{Repko}
\frac{\phi}{l}=\frac{112\pi G_F\alpha_{em}}{45\sqrt{2}}\left[\ln \left(\frac{m_W}{m_e}\right)^2 - \frac{8}{3}\right]\frac{\omega^2T_{\nu}^2}{m_W^4}(n_{\nu} - n_{\bar{\nu}}).
\end{equation}
This gives for the dimensionless neutrino chemical potential $\mid\xi_{\nu}\mid=\mid\mu_{\nu}\mid/T=0.01$ at the present relic neutrino temperature $T_{\nu}\sim 2~{\rm K}$ and for EHE photons $\omega=10^{20}~{\rm eV}$ only $\phi\sim 4\times 10^{-16}~{\rm rad}$ at the horizon size $l=l_H=H_0^{-1}=4.3\times 10^3~{\rm Mpc}$. For radio waves or even for optical photons $\omega\sim O({\rm eV})$ such rotary power would be scanty at all. 

Another result for extragalactic sources can be obtained for the neutrino-antineutrino sea embedded into isotropic plasma where transversal photons get the effective mass $\omega^2- k^2 =\omega_p^2\neq 0$ \cite{Mohanty:1997mr}:
\begin{equation}\label{Mohanti}
\frac{\phi}{l}=\frac{G_F\alpha_{em}}{3\sqrt{2}\pi}\left(\frac{\omega_p^2}{m_e^2}\right)(n_{\nu} - n_{\bar{\nu}}).
\end{equation}
First, this result does not depend on a photon frequency as well as in the case (\ref{angle}) discussed below. For the plasma frequency $\omega_p=5.65\times 10^4\sqrt{n_e}~sec^{-1}=6.5\times 10^{-13}~{\rm eV}$ given by a small $n_e\sim 3\times 10^{-4}~cm^{-3}$ in the galo of a Milky-Way-like galaxy \cite{Hammond:2012pn} one obtains at the distance $l=l_H$ a very small value of the optical rotation $\mid\phi\mid\simeq 8\times 10^{-44}~{\rm rad}$. For the intergalactic electron density $n_e\sim 10^{-5}~cm^{-3}$ this rotation angle would be lowered by the factor 30. Authors \cite{Mohanty:1997mr} overestimated rotary power (\ref{Mohanti}) getting $\phi\sim 10^{-36}$ instead when they substituted $n_e=3\times 10^{-2}~cm^{-3}$ and used an archaic demand that the neutrino energy density should not exceed the closure density of the universe. This led them to a big neutrino asymmetry density exceeding significantly the value $\mid n_{\nu}-n_{\bar{\nu}}\mid=0.01T^3_{\nu}/6$ we have just used for the same $\mid\xi_{\nu}\mid=0.01$ in Eq. (\ref{Mohanti}).

Now let us turn to the analysis of the main result (\ref{angle}).
Substituting for $\Pi_2(0)$ its explicit form (\ref{polarization}) taken from Eq. (B1) in \cite{Boyarsky:2012ex} one obtains from Eq. (\ref{angle}) the result independent of the wave frequency,
\begin{eqnarray}\label{shapo}
\frac{\mid\phi\mid}{l}=&&\mid\Pi_2(0)\mid =\frac{4\alpha_{em}G_F}{2\pi\sqrt{2}}\left(\frac{2}{9}\right)\mid\sum_a\Delta n_{\nu_a}\mid=\nonumber\\&&= \frac{4\alpha_{em}G_F}{9\pi\sqrt{2}}(0.244 T_{\gamma}^3)\mid\eta_{\nu}\mid .
\end{eqnarray} 
This gives $\mid \phi\mid=1.23\times 10^{-6}~{\rm rad}$ at the distance $l=l_H$ where we used the upper limit from WMAP+He bounds on the total neutrino asymmetry $\eta_{\nu}=\sum_a\eta_{\nu_a}$,  $a=e,\mu,\tau$, obtained in \cite{Castorina:2012md,Mangano:2011ip}: $-0.071<\eta_{\nu} < 0.054$ . These bounds  define the total relic neutrino-antineutrino asymmetry density
$\sum_a\Delta n_{\nu_a}=n_{\gamma}\eta_{\nu} =(0.244T_{\gamma}^3)\eta_{\nu} $ entering the lepton number density in $\Pi_2(0)$. When flavours equilibrate due to oscillations before BBN in the presence of a non-zero mixing angle (for the case $\sin^2\theta_{13}=0.04$ used in \cite{Castorina:2012md,Mangano:2011ip} involving all active neutrinos) the total neutrino asymmetry is distributed almost equally among the different flavors, leading to a final asymmetry, $\eta_{\nu_e}^{fin}\approx\eta_{\nu_x}^{fin}\approx \eta_{\nu}/3$, where $x=\mu, \tau$. We remind the definition of the asymmetry given by the partial degeneracy parameter $\xi_{\nu_a}\equiv \mu_{\nu_a}/T$,
$$
\eta_{\nu_a}\equiv \frac{n_{\nu_a} - n_{\bar{\nu}_a}}{n_{\gamma}}=\frac{1}{12\zeta(3)}\left[\pi^2\xi_a + \xi^3_a\right],
$$
which accounting for $n_{\gamma}=2\zeta(3)T^3/\pi^2=0.244T^3$ gives for $\xi_{\nu_a}\ll 1$ the well-known fermion asymmetry density $\Delta n_{\nu_a}\equiv n_{\nu_a} - n_{\bar{\nu}_a}\simeq\xi_{\nu_a}T^3/6$. Note that the electron density in intergalactic region $n_e$ is much less than the relic neutrino density at present, $n_e\ll \langle n_{\nu_a,\bar{\nu}_a}\rangle = 112~cm^{-3}$. This is a reason why we neglected electron contribution to the total lepton number ${\rm L}_{tot}$ in polarization operator $\Pi_2(0)$. Note that accounting for the gravitational clustering of relic neutrinos in cold dark matter halos one can expect an increase of the rotation angle given by Eq. (\ref{shapo}) due to a non-relativistic neutrino (antineutrino) overdensity $n_{\nu, \bar{\nu}}/\langle n_{\nu, \bar{\nu}}\rangle\sim 10\div 100$ \cite{Singh:2002de} obeying some upper bounds on the sum of neutrino masses.

The Faraday rotation measure (RM) for the competing rotation of polarization vector  at the angle $\phi=\lambda^2{\rm RM}$ in an intergalactic magnetic field (IGMF) $B_{\parallel}/\mu G $, $B_{\parallel}=({\bf B}\cdot {\bf k})/k$, is given by the well-known formula \cite{1983flma....3.....Z}  :
\begin{equation}\label{Faradayrotation}
{\rm RM}=0.81\times 4.3\times 10^{9}\left(\frac{n_e}{cm^{-3}}\right)\left(\frac{B_{\parallel}}{\mu G}\right)\left(\frac{l}{l_H}\right)\frac{{\rm rad}}{m^2},
\end{equation}
where we substituted $l_H=4.3\times 10^9~pc$. Together with the estimate of the electron density in intergalactic medium $n_e\simeq 3\times 10^{-4}~cm^{-3}$ and upper bound on IGMF $(B/\mu G) < 10^{-3}$ known from CMB observations  this leads to the upper bound on RM:
\begin{equation}\label{upper}
\mid{\rm RM}\mid< 10^3\left(\frac{l}{l_H}\right)~\frac{{\rm rad}}{m^2},
\end{equation} 
that corresponds to observable values ${\rm RM}=\pm 10\div \pm 100~{\rm rad}/m^2$ for quasars in the radio wave band $\lambda\sim 1~m$. For a microwave $\lambda\sim {\rm O}(cm)$ Eq. (\ref{upper}) gives the bound on the angle $\phi< 0.1(l/l_H)~{\rm rad}$.

Of course, somewhere in optical band $\lambda\geq 10^{-5}/2~cm$ ($\omega\leq O(~{\rm eV})$) the corresponding rotary power yields to 
our result in Eq. (\ref{shapo}).

In the paper \cite{Neronov:1900zz} authors reported the lower bound on IGMF $B\geq 3\times 10^{-16}~G$ which stems from the nonobservation of GeV gamma-ray emission from electromagnetic cascade initiated by $\gamma$-ray flux from blazars with energy $E_{\gamma}\sim 10^{12}~eV$ in extragalactic medium.
The corresponding lower bound on RM turns out to be:
\begin{equation}\label{lower}
\mid{\rm RM}\mid>3\times 10^{-4}\left(\frac{l}{l_H}\right)~\frac{{\rm rad}}{m^2}.
\end{equation} 
For a microwave $\lambda\sim {\rm O}(cm)$ Eq. (\ref{lower}) corresponds to the lower bound on the rotation angle $\phi> 3\times 10^{-8}(l/l_H)~{\rm rad}$ for which the birefringence effect(\ref{shapo}) becomes more visible. 

Nevertheless, there are no instruments at present to measure such tiny effects in astrophysics.

\end{document}